# Simulating an Object-Oriented Financial System in a Functional Language[*]

Lee Braine, Keith Haviland,
Owen Smith-Jaynes, Andy Vautier

Andersen Consulting
2 Arundel Street
London, U.K.

Chris Clack

Department of Computer Science
University College London
Gower Street
London, U.K.

**Abstract.** This paper summarises a successful application of functional programming within a commercial environment. We report on experience at Andersen Consulting's London Solution Centre with simulating an object-oriented financial system in order to assist analysis and design. The work was part of a large IT project for an international investment bank and provides a pragmatic case study.

## 1 Introduction

Functional languages such as Miranda[†] [Tur85], Standard ML [MTH90], Haskell [PHA97] and Clean [Pla95] are used extensively in academia for research and teaching. These languages offer a number of well-known software engineering and formal methods benefits, including rapid development, clear and concise expression of algorithms, and complete type safety. However, despite these benefits, functional languages have experienced only limited acceptance by industry – mainly due to historical problems (such as inefficient compilers and limited input/output) that have been largely extenuated by research in the past decade.

This paper summarises a recent successful application of functional programming within a large IT project (over 100 developers). The project was undertaken by Andersen Consulting's London Solution Centre in partnership with an international investment bank and included the construction of a large financial software system using object-oriented and component-based techniques. A functional language was employed in the analysis and design stages to specify complex algorithms precisely and to simulate the high-level behaviour of the entire system. Following successful simulation, the specifications provided validated algorithm designs for subsequent implementation in C++. In this paper, we highlight the benefits which resulted from the functional simulation and also identify aspects that proved difficult to simulate.

## 2 Related Work

Simulation techniques are used extensively to model complex system behaviour. A variety of languages and tools are applied to problems ranging from analysing the movement of data packets in a network to predicting the value of financial derivatives – see [LK98] for an introduction to simulation modelling and its applications.

Functional programming has been applied to a wide range of problems ([RW95] summarises some recent applications). Illustrative examples of its use in simulation include:

- [PW93] and [Poh94] discuss simulation programming in a functional language;
- [PM93] presents a case study of Amoco's use of Miranda to simulate oil reservoirs;
- [EL83] describes a high-speed digital simulation using a functional language approach;
- [GSW93] presents a functional programming solution to a computational fluid dynamics problem.

---

[*] This document is based on the paper "Simulating an Object-Oriented Financial System in a Functional Language", L. Braine and C. Clack, in *Proceedings of the 10th International Workshop on Implementation of Functional Languages (IFL'98)*, pages 487-496, London, September 1998.
[†] Miranda is a trademark of Research Software Ltd.

Functional languages are also used in the closely-related areas of prototyping and specification; examples include prototyping computer-aided design applications [DB85], specifying image processing primitives [PC94], specifying complex tree transformations [Hec88] and describing telecommunications systems [Dem89].

However, there is little in the literature on commercial applications of functional languages to simulate financial systems. One reason for this is the often-held view that real-world processes can be represented more naturally in simulations using an object-oriented approach, as first advocated by SIMULA [DN66], than using a standard functional approach. Subsequent object-oriented languages, such as Smalltalk [GR83] and Eiffel [Mey91], widened this gap as functional languages offered few object-oriented features – see [BC96] for a brief history of research into object-oriented functional programming.

## 3  The Application

The subject of simulation activities was a large financial system within an international investment bank. The system contained a set of complex business processes requiring novel optimisation and approximation algorithms in order to perform effectively. Full details are both proprietary and confidential but, for the purposes of this paper, we present a brief overview in this section. At the highest level, the application involves partitioning a collection of elements into two sets. The problem is to maximise an aggregate property of the population contained in one of the sets by deciding which elements can be a member by satisfying a number of business constraints.

Solving this problem required simulation of two key modes of processing that correspond closely to (i) real-time processing and (ii) batch processing. For real-time processing, standard discrete-event simulation techniques were employed, such as annotating data items with explicit time tags. For batch processing, the approach was to rapidly prototype the algorithms in a functional language, including simulation of some object-oriented aspects and system functionality.

The application contained a number of key data representations:

- Elements were categorised into a partially-ordered list of queues, with each queue itself a partial order.
- Queues were grouped according to common business criteria and distinct processing rules were applied to each group depending upon reference data.
- Data elements contained many cross-references to data items in other queues. These dependencies introduced possible deadlocks into the system by requiring application of potentially conflicting business constraints. This forced the construction of super-groupings of data-dependent items which spanned queues.
- By combining the ordering inherent in the list of queues and the key business constraints, the data elements can be connected as a graph. Various graph-theoretic algorithms (such as Tarjan's algorithm identifying the strongly connected components of a graph) can then be used to manipulate the data.

The goals for this simulation work were to:
1. rapidly develop and evaluate complex algorithms;
2. validate interactions between system components;
3. specify algorithms in an object-oriented style (for subsequent implementation in C++).

## 4  Simulation Language

We now consider the rationale behind simulating in a functional language to meet the goals identified in the previous section. The first goal favours the use of a functional language, but the last two goals favour a standard object-oriented language. In particular, the project methodology was object-oriented and employed a component-based approach, requiring modelling of the system's object-oriented algorithms and components.

Simply prototyping the algorithms in the final implementation language, C++, was not a viable option because it would have taken too long to develop underlying components before the high-level algorithms could be simulated. We decided to use a functional programming language rather than an object-oriented rapid application development language, such as Smalltalk, for the following reasons:

- The speed and clarity with which the algorithms could be expressed and validated was the most important consideration. The language features offered by functional programming (e.g. higher-order functions, lazy evaluation and complete type safety) provide benefits of clarity, conciseness and speed of expression in excess of many imperative languages. See [PM93] for illustrative metrics.
- The only parts of the system design that required precise simulation of their object-oriented aspects were the high-level algorithms and their method calls via component interfaces. The internals of all other components could be simulated using a purely functional approach to speed development. Additionally, the extra work involved in coercing a functional language to simulate the object-oriented aspects (e.g. inheritance hierarchy, dynamic despatch and mutable state) could be minimised by applying techniques from recent object-oriented functional programming research.
- Andersen Consulting's London Solution Centre has rapid application development expertise using functional programming and could utilise its academic links to obtain specialist input as required (e.g. functional programming simulation techniques).
- The execution speed of the simulation was not important, so the relatively slow speed of functional languages was not a disadvantage.

Miranda was the functional language of choice, mainly because it is commercially supported and provides an interpreted environment to assist rapid application development. If an interpreted environment were not as important, a compiled functional language such as Clean could have been used instead.

## 5  Simulation Environment

Simulation activities were performed by the Simulation Team over a period of 6 months. The software environment was the interpreted functional language Miranda running on the Solaris operating system. Multiple concurrent environments were executed on a SUN Enterprise 4000 server (eight UltraSPARC CPUs with 4GB shared RAM) accessed from networked PCs running Windows NT. Individual environments used between 50MB and 1GB RAM, depending on the complexity of the simulation.

## 6  Simulation Methods

Different levels of simulation fidelity were required for different parts of the system. For example, in order to provide algorithm specifications for subsequent implementation in C++, it was necessary to simulate those object-oriented algorithms precisely. Additionally, the use of component interfaces had to be simulated precisely in order to validate interactions between system components. However, only the behaviour (not the internal details) of those components had to be simulated, permitting the use of statistical approximation techniques in some cases.

There are important semantic differences between the object-oriented and functional paradigms – [BC96] overviews the main theoretical differences. Some of the obvious mappings are sufficient (e.g. function signatures for modelling component interfaces), but others are insufficient (e.g. abstract data types for modelling classes). In order to simulate object-oriented designs using a purely functional language, it is necessary to resolve these semantic differences by applying techniques from object-oriented functional programming research.

The remainder of this section discusses the three key simulation methods that were used.

### 6.1  Real-Time Simulation at the Component Level

At the most abstract level, the application consists of a number of concurrent communicating components. The requirement was to model the real-time behaviour of these components, specifically the timing of data interchange between the components and the operation of the internal algorithms according to the data arrival times.

At this level, there was no requirement to model a complex class hierarchy (for example, components do not exhibit inheritance characteristics). Each component was concisely expressed as a lazily-evaluated "spinning function" – that is, a function which takes one or more infinite streams of data as input and produces one or more infinite streams of data as output (collected into a tuple). An additional accumulating parameter was used to hold the local state for each component.

The input and output streams modelled the independent, buffered, communication channels between the components; each stream was implemented as a lazy list of time-tagged data items. Network starvation (and possible deadlock due to blocking reads of the input streams) was avoided through the use of *hiatons* [Sto85]; whenever a component has nothing to output on a stream, it explicitly outputs an empty data item together with an appropriate time tag.

### 6.2 Behavioural Simulation of Component Internals

For many components, it was only necessary to model the component behaviour and a straightforward style of functional programming was used. It was observed that the use of functional programming features such as higher-order functions led to a great reduction in code size; this was particularly beneficial when developing complex algorithms as the functional notation provided a very concise and understandable specification. Furthermore, Miranda's interpreted environment and static type inference strongly supported an exploratory programming style; the benefits were minimal build time and very little run-time debugging.

The functionality of some parts of this code was determined by the results of statistical approximation techniques. This enabled several parameters, such as event frequency, to be set explicitly and the resultant effects observed by executing the simulation. Miranda's interpreted environment also provided a rapid turnaround when performing investigations of these parameters.

### 6.3 Specifying Object-Oriented Algorithms

The most complex simulation task was to simulate the action of some algorithms at a sufficiently detailed level that the functional code could be used as an object-oriented specification of the algorithm to be subsequently implemented in C++. This required simulation of several object-oriented features that are not provided by Miranda.

In this case, the coding style adopted by the Simulation Team was similar to the style of target code produced by the CLOVER compiler [BC96,BC97a], particularly in the areas of simulating classes with inheritance, overloading, overriding, and dynamic despatch. Full details can be found in the research literature, but we rehearse the essence of the technique here:

- *Classes, inheritance and the meta type system*

    In order to simulate certain aspects of the object-oriented system, it was necessary to create a meta type system. For example, simulating subsumption (the notion of manipulating an object which could be one of several possible subtypes) required inheritance hierarchies to be flattened by creating a new type for each hierarchy that represented the root superclass, and then encoding each subclass as an alternative in an algebraic type.

- *Dynamic method despatch*

    The object-oriented notion of dynamic method despatch conflicts fundamentally with the functional notion of static type safety. However, by using the meta type system mentioned above, efficient method despatchers were constructed that simulated dynamic despatch. These simply pattern-matched on the type constructor to dynamically select the appropriate method.

- *Simulating assignment*

    In order to provide detailed object-oriented specifications, it was necessary to simulate assignment. This included creating multiple single-assignment identifiers in order to represent multiple assignments to a single object. It should be noted, however, that recent functional

research can ameliorate such a plethora of identifiers by applying suitable lexical scoping rules, for example the Clean language allows identifiers on the right-hand-side of an expression to be reused on the left-hand-side – they are then internally tagged with a number by the compiler (see [AP97] for further details).

# 7 Results

During the process of simulation, several algorithms were:

1. developed using Miranda;
2. validated through a high-level Miranda simulation of the entire system;
3. used as specifications for subsequent implementation in C++.

These activities proved highly successful and the key results are reported below:

- *Rapid development*
  Miranda code was produced far more rapidly than C++ code with similar functionality (this has been estimated as a factor of approximately 5 times). We expect that the productivity gain was mainly due to two key factors: (i) the often-expressed desirable language features of functional programming (higher-order functions, lazy evaluation, automatic memory management, etc.), and (ii) the traditional benefits of an interpreted environment (negligible build time, rapid turnaround cycle, interactive testing of individual functions, etc.).
- *Concise expression*
  Miranda specifications were substantially more concise than C++ code with similar functionality. As an illustrative example, a key algorithm expressible in 6 pages of Miranda code translated into approximately 20 pages of C++ code. The conciseness of functional programs is often reported in the research literature and as a result of scientific applications; here it has been validated for a financial application.
- *Simulation as executable specification*
  By employing a functional language at the analysis and design stages, complex processes were simulated in advance, allowing designs to be optimised early in the project lifecycle. The functional programs also served as *executable specifications* [Tur85a] – the algorithms were tested on actual data and, once confident of correctness, used as validated specifications for the final designs. Unit testing of the actual system found that fewer errors existed in C++ code that had been first specified and simulated using Miranda. Considering that this included many of the most complex algorithms in the system, we believe that this result is one of the most important justifications for using simulation work early in the lifecycle of a complex project.
- *Limitations*
  Simulations required extensive computing resources. The most notable was a very large memory space – up to 1GB of heap space in some simulations. The reasons are both application-specific (e.g. the requirement for full and complete traces of events, used later for analyses into the operational behaviour of the algorithms) and functional language-specific (e.g. the maintenance of closures during lazy evaluation). Although execution time of simulations was not important, many took 5 to 10 minutes, with the most complex taking approximately 3 hours.

## 8 Further Work

There is potential for further application of simulation and specification techniques within this current project and we are also liaising with Andersen Consulting's "Centre for Process Simulation and Modeling" on the opportunity for novel simulation techniques in other projects.

We have also identified several ways in which current functional language implementations could be improved in order to support simulation and specification activities for financial applications:

- Our experience indicates that functional programming would benefit from the incorporation of more object-oriented features. Not only do general simulation activities benefit naturally from an object-oriented approach but, with many actual software systems being built using component/object-oriented techniques, simulations can more accurately specify the actual system. However, incorporating object-oriented features into functional programming is not trivial and should be regarded as on-going research.

- On a large project, many specifications are delivered as part of a set of documents created using a standard wordprocessor, such as Word or WordPerfect. This necessitates cutting-and-pasting executed specifications into project documentation; the procedure is potentially error-prone and could be resolved by executing the wordprocessing documents themselves. This "literate script" programming style is available in Miranda for the LaTeX typesetting system; we would like to see other languages adopt this style and extend it to Word as well as LaTeX.

- A link with object-oriented design tools, such as Rational Rose, could allow preliminary high-level designs to be generated automatically from validated simulation code. This may require code annotations to guide the generator, but would be particularly useful for creating top-level components and component interface diagrams from simulation classes.

- Many simulation tools permit process execution to be visualised through the use of graphical animation. This allows the behaviour of complex algorithms to be understood by a wider audience and can often provide additional insights (such as identifying bottlenecks). Animation could be added to functional simulations by extending state classes to capture event information and output it for post-simulation animation.

- We would like to explore the use of other functional languages in order to benefit from both additional language features and from greater execution speed. However, our experience indicates that for rapid simulation and specification purposes one of the most useful system features is an interpretive environment. We urge functional language implementors to provide both interpreted and compiled modes of execution.

    Additionally, maintaining simulation traces (for later analysis) required the maintenance of very large data structures; unfortunately, the Miranda heap limit of approximately 1GB (despite 4GB of RAM being available to us) precluded some particularly complex simulations. With the high-end hardware now available, we urge system implementors to remove such arbitrary limits.

## 9 Summary and Conclusion

In this paper, we have presented a successful application of functional programming to the simulation and specification of an object-oriented financial system. A variety of simulation methods were employed to model the different parts of the system at appropriate levels of detail. This included simulation of a real-time system at the component level, behavioural simulation of component internals and detailed specification of core object-oriented algorithms.

The results were highly successful and highlighted the key benefits of using a functional language. These included very rapid development (we estimate 5 times faster than prototyping in C++), concise expression (a validated design in 6 pages of Miranda translated to about 20 pages of C++ code) and use as an executable specification which can be tested with real data. Subsequent phases of the project demonstrated that a functional language can serve as a design language for the object-oriented implementation, with the validated Miranda designs resulting in high quality C++ code.

In conclusion, this project has demonstrated the great worth of including simulation activities early in the lifecycle of a complex financial project and, in particular, identified some key benefits obtained by specifying and simulating in a functional language.

## Acknowledgements

We wish to acknowledge the contributions made during the course of this work by Tony Wicks (of SearchSpace Ltd.) to the data analysis.